\documentclass[a4paper, amsfonts, amssymb, amsmath, reprint, showkeys, nofootinbib, twoside, floatfix]{revtex4-1}
\usepackage[english]{babel}
\usepackage[utf8]{inputenc}
\usepackage[colorinlistoftodos, color=green!40, prependcaption]{todonotes}

\usepackage{adjustbox}
\usepackage{placeins}
\usepackage{bm}
\usepackage{xcolor}
\usepackage{amsthm}
\usepackage{mathtools}
\usepackage{physics}
\usepackage{xcolor}
\usepackage{graphicx}
\usepackage[left=23mm,right=13mm,top=35mm,columnsep=15pt]{geometry} 
\usepackage{adjustbox}
\usepackage{placeins}
\usepackage[T1]{fontenc}
\usepackage{lipsum}
\usepackage{csquotes}
\usepackage[pdftex,
    colorlinks=true,
    linkcolor=black,
    citecolor=black,
    urlcolor=blue,
    pdftitle={Article},
    pdfauthor={Author}
]{hyperref} % For hyperlinks in the PDF
\begin{document}
\title{Nano-silica based Aqueous Colloidal Gels as Eco-friendly Thixotropic Lubricant}

\author{Arun Kumar,\textit{$^{a\dag}$}  Vivek Kumar,\textit{$^{b\dag}$} Yogesh M. Joshi,\textit{$^{b*}$} Manjesh Kumar Singh\textit{$^{a}$}}
    \email[Correspondence email address: ]{joshi@iitk.ac.in; manjesh@iitk.ac.in}
    \thanks{$\dag$ Equal contribution}
    \affiliation{$^{a}$~Department of Mechanical Engineering, Indian Institute of Technology Kanpur, Kanpur-208016, Uttar Pradesh, India\\
$^{b}$~Department of Chemical Engineering, Indian Institute of Technology Kanpur, Kanpur-208016, Uttar Pradesh, India}

% \date{\today} % Leave empty to omit a date

\begin{abstract}
The environmental risks posed by traditional oil- and grease-based lubricants can be significantly mitigated by adopting water-based alternatives engineered with superior rheological performance. In this work, we present a fundamentally new and environmentally sustainable aqueous thixotropic colloidal gel of silica nanoparticles formed in the presence of NaCl. We conducted a systematic and detailed investigation of their rheological and tribological characteristics. The tribological performance was evaluated against dry and water-lubricated conditions for steel–steel interface. Our experiments demonstrate that the tribological performance of the formulated nanoparticle gel can be optimized by tuning its rheological properties. A combination of super-low friction and negligible wear was observed. The friction coefficient reduced by up to 97.46\% (from 0.63 to 0.016) compared to dry sliding, and by 97.04\% (from 0.541 to 0.016) compared to water lubrication. Similarly, the specific wear rate decreased by up to 99.62\% and 96.10\% under under dry conditions and water lubrication respectively. This performance is attributed to a thixotropic, chemically robust gel formed via van der Waals interactions between silica flocs, enabling self-repairing properties, continuous tribo-film formation, and a nano-bearing effect from silica nanoparticles. These attributes enable the gel to maintain and regain its structure during periods of non-shear while also forming a thin film with sufficiently low viscosity to slip into the interfacial contact zone and continuously replenish it with lubricant.
\end{abstract}

\keywords{friction, wear, nano-silica, rheology, aqueous lubrication, tribology, thixotropy}

\maketitle
\section{Introduction}
Conventional oil and grease lubricants pose environmental and human health risks attributed to their toxic additives \cite{Boyde2002GreenLubrication,Singh2022AqueousLubrication,Madanhire2016LubricantEnvironment}. Consequently, there has been a transition towards water-based lubricants that prioritize environmental and economic considerations \cite{Li2021SurfaceFriction,  Adibnia2021PhytoglycogenSuperlubricants, Rosenhek-Goldian2018TrappedSurfaces, Singh2015PolymerExperiments,Singh2020,manjesh.macromol,Nalam2013ExploringSolvents}. However, water's inherent properties limit its practical use as an effective lubricant. Water's low viscosity typically yields lubrication film thicknesses approximately two to three orders of magnitude smaller than conventional oils under hydrodynamic lubrication conditions \cite{Kim2015WaterCoating,Wang2012InfluenceLubrication}. Moreover, under high contact pressure, water gets squeezed out of the interface owing to its low pressure-viscosity coefficient, resulting in direct surface-to-surface contact.

 Simply augmenting viscosity does not ensure effective lubrication. Therefore, there is a necessity to utilize non-Newtonian lubricants \cite{Veltkamp2023LubricationFluids,Ortigosa-Moya2019SoftFluids,Fernandez2013MicroscopicSuspensions,Yang2016ASimulation,Hsu2021ExploringParticles}, capable of dynamically adjusting viscosity to accommodate fluctuating operating conditions, thus enhancing overall performance. Issues of shear jamming in shear thickening fluids at a critical shear \cite{Fernandez2013MicroscopicSuspensions,Hsu2021ExploringParticles} directs us to consider thixotropic lubricants that change their rheological properties according to the applied deformation field (shear flow). These lubricants should be capable to form a thin film of sufficiently low viscosity to reduce friction and wear of surfaces in motion. Under the influence of a deformation field, the microstructure of an existing colloidal gel gradually breaks, leading to a decrease in viscosity. Upon cessation of the flow, the microstructure gradually builds-up over time, resulting in an increase in viscosity \cite{Bhattacharyya2023OnMaterials}. In a recent investigation, we demonstrated the effectiveness of thixotropic colloidal gel, of 2D Laponite clay nanoparticles having anisotropic charge distribution on edges and faces, as a lubricant \cite{Kumar2024TribologicalNanoparticles}. We observed a reduction in the coefficient of friction (CoF) to a certain degree. However, the instability of the CoF curve under fully submerged lubrication constrains its practical application. Additionally, achieving exceptionally low friction under high contact pressure and low speeds is still a challenge. In recent years, additives \cite{Chen2022Surfactant-ModifiedAnticorrosion,Li2021InfluenceCeramics,Radice2006EffectSuspensions,Ding2018TheAdditives,Huang2019SynergisticAdditives,Cui2020RoleCeramics,Zhang2014PreparationAdditive,Zhang2012PreparationAdditive,Wang2016LayeredAdditives} have been used to develop water lubrication systems for numerous industries including manufacturing, automotive, ship building, food and biomedical \cite{Kim2015WaterCoating}.

With their nanoscale dimensions and surface effects, nanoparticles serve as promising lubricant additives \cite{Soltannia2023EnhancedContactsb,Saidi2020ImprovedAdditives,Erdemir2016Carbon-basedOils,Chen2020UltrastableNanoparticles,Xie2016LubricationContacts,Jiao2011TheAdditives,Li2006TribochemistryAdditives,SalinasRuiz2021InterplayLubricants}. Their multifaceted mechanisms of action include hydration film formation, the ball-bearing effect, mending of surface cavities, and the development of protective films through either physical adsorption or through complex tribochemical reactions at the sliding interface \cite{Mou2022One-stepDurability,Yang2023BiomimeticPerformances, Xie2023Calcitriol-LoadedTreatment, Liu2023Meniscus-InspiredExercise, Adibnia2021PhytoglycogenSuperlubricants, Wang2024FabricationPerformance}. Such mechanisms impart both anti-friction and anti-wear properties.

Silica (SiO\textsubscript{2}) nanoparticles owing to their cost-effective and environment-friendly nature, garner significant attention among other nanoparticles \cite{Li2021InfluenceCeramics,Ding2018TheAdditives,Lin2019ExcellentConditions}. SiO\textsubscript{2}, often utilized in pair with other nanoparticles as a modifier, help in enhancing the tribological properties of the tribo-system \cite{Wang2020NovelProperties,Zhang2012PreparationAdditive,Liu2020PreparationAdditive}. For instance, in one such study carbon nanotube (CNT) and spherical SiO\textsubscript{2} nanoparticles as lubricant additives in deionized water demonstrate reduction of both the friction coefficient and wear volume. This is credited to the sliding-induced tribofilm of CNTs and the potential ball-bearing effect of nano-SiO\textsubscript{2} \cite{Xie2021TribologicalAlloy}. Combined effect of SiO\textsubscript{2} and 2D material such as graphene oxide (GO) was studied by Guo \emph{et al.} \cite{Guo2018EnhancedAdditive}, where it was observed that the combination of the mechanical properties of GO nanosheets and the microscopic ball-bearing effect of silica nanoparticles contributes to the lubricating and anti-wear properties of aminated silica-modified GO in water. Experimental studies and simulations demonstrate that SiO\textsubscript{2} nanoparticles have promising lubrication capabilities for ceramics also \cite{Wen2016AtomicField, Li2021InfluenceCeramics, Lin2019ExcellentConditions, Ding2018TheAdditives}. 

Colloidal silica, suspensions comprising fine, amorphous, nonporous, and typically spherical silica particles in a liquid phase, are utilized in the production of nanotextured surfaces for micro/nano electromechanical systems via spin-coating. Surfaces generated through this method have demonstrated favorable tribological properties concerning adhesion and friction \cite{Zou2006NanotribologySurface,Zou2006AdhesionSolution}. When exposed to accelerators such as NaCl, KCl, CaCl\textsubscript{2}, colloidal silica can undergo gel formation, with the process and rate of gelation being controllable \cite{Sogaard2018SilicaAnalysis}. 

The literature highlights the tribological potential of SiO\textsubscript{2} nanoparticles. In an area dominated by non-biodegradable synthetic oil-based gels \cite{Wang2024FabricationPerformance,Wang2024ImprovedLubricant}, aqueous nanofluids and multi-nanoparticle blends, this work introduces a simple, scalable, and eco-conscious lubrication strategy based solely on colloidal nano-silica. Addressing the limitations of a general class of water-based lubricants between metallic contacts particularly under high load conditions \cite{Jia2019ALubricant}, this study reveals the untapped potential of silica gels as robust aqueous lubricants by exploiting their thixotropic nature. Through a systematic investigation of their rheological and tribological behavior, we establish a clear link of structural recovery with friction and wear reduction, paving the way for sustainable, high-performance thixotropic gel-based lubrication.

\section{Experimental}
\subsection{Materials and Chemicals}

Stock suspension of spherical silica colloidal nanoparticles (Ludox\textsuperscript{\textregistered} TM-40, 40 wt\% suspension in H\textsubscript{2}O) with diameter in the range of 20 to 22 nm, and specific surface area $\approx$ 140 m\textsuperscript{2}/g, was procured from Sigma-Aldrich. The salt (NaCl) was procured from LOBA CHEMIE). Tribological tests were conducted on AISI 304 austenitic stainless-steel discs with a Rockwell hardness of 25 HRC. The discs (15 $\times$ 15 $\times$  3 mm) were precision-cut using wire electro-discharge machining. Subsequently, they underwent a polishing process, starting with SiC emery paper from P320 to P5000 grit, followed by cloth polishing using diamond lapping compounds with particle sizes of 6, 3 and 1$\mu$m. Ultrasonic cleaning with acetone, followed by de-ionized water, was performed to eliminate any residue of grime, oil, or fatty acids. The discs were then blow-dried. Before each experiment, the discs were wiped with ethanol to eliminate any potential contaminants that could impact the material's tribological properties.

\subsection{Rheological Tests}

All the rheological experiments of this study have been done on the stress-controlled Anton Paar MCR-501 Rheometer at 25$^\circ$C. To avoid the errors in the measured data due to slip, the sand-blasted concentric cylinder geometry with a 26.65 mm diameter bob and 28.915 mm I.D. cup was used. To prevent the drying of the sample surface during the measurement, the top surface of the sample was covered with low-density silicon oil. Since the silica gel systems are non-ergodic in nature and consequently their microstructure evolves with time, all the gel-forming systems were initially shear-melted at a high shear rate of 300 s\textsuperscript{-1} to erase the memory of previous deformation fields. Subsequent to shear melting, systems were subjected to shear rate sweep experiments in the direction of high to low shear to determine the flow curve. In addition to the flow sweeps, the stable silica gel systems were also subjected to cyclic frequency sweep experiments at $\omega\in{[1,10]}$ rad/s in the linear deformation domain for 1 hour to understand their aging behavior. Furthermore, to determine the effect of aging time on the relaxation dynamics of the systems, after shear melting, the stable systems were first allowed to age for different aging times ($t_w$), and subsequently, the creep deformation field was applied. During the waiting period ($t_w$), the systems were subjected to a very small deformation field of  0.1 \% strain amplitude and 0.63 rad/s angular frequency.

\subsection{Tribological Tests}

Sliding tests were conducted using a uni-directional rotary pin-on-disc tribometer (Magnum Engineers, Bangalore) in a ball-on-disc configuration (schematic shown in Figure S1, Section S1: Supporting Information (SI)). This configuration is commonly used to test semi-solid lubricants such as grease, as it can replicate the contact pressures experienced in these systems \cite{Cyriac2016EffectSpeeds,Merieux2000Shear-degradationLubrication}.  The frictional force load sensor was accurately calibrated prior to testing. All experiments utilized self-mated AISI 304/AISI 304 tribopairs, a commonly utilized steel-steel combination found in various industries and is known for its resistance to oxidative corrosion. To ensure consistent contact between the disc and ball, proper alignment was maintained for each test. The experiments were repeated thrice under the following conditions: normal load = 5 N (mean Hertzian contact pressure of 730 MPa), linear sliding speed = 0.04 m/s, temperature = 23$\pm$0.3$^\circ$C, relative humidity = 59$\pm$3\%, and a test duration of 20 min. The test parameters were selected following preliminary experiments covering a range of loads from 1 to 10 N and speeds from 0.005 to 0.3 m /s (refer to Figure S2-S4, Section S2: SI). These preliminary tests indicated that the lubricant performed optimally at a combination of 5 N load and 0.04 m/s speed, achieving an adequate level of wear and friction within a reasonable duration. The maximum average surface roughness (R\textsubscript{a}) and diameter of the balls were 0.20 $\mu$m and 6 mm, respectively. Each disc surface was uniformly covered with a fixed mass of 1 g of lubricant for every test. The initial conditions of the tribological tests were carefully controlled for consistency across all experiments. 

\subsection{Basic Surface Analysis}

The surface roughness, and wear profile of the AISI 304 disc were evaluated using a non-contact 3D optical profilometer equipped with white light interferometry (Contour GT-K, Bruker). Wear volume and hence specific wear rate (SWR (mm\textsuperscript{3}/N-m) were calculated to provide insights into material loss. Wear scars on the counter surface (AISI 304 balls) were examined using an optical microscope (OLYMPUS BX51M). The worn disc surfaces were analyzed without any conductive coating using a Scanning Electron Microscope (SEM) (CARL ZEISS EVO 50). Chemical composition and quantitative analyses were conducted using Energy Dispersive X-ray Spectroscopy (EDS) by Oxford Instruments, utilizing the INCA and Aztec software.

\subsection{Raman Spectroscopy}

Raman spectra were obtained utilizing a Princeton Instruments Acton Spectra Pro 2500i spectrometer equipped with a Diode-pumped solid-state laser (532 nm) delivering an initial power of 40 mW. Spectra were acquired employing a 1024$\times$256-pixel charge-coupled device (CCD) detector in a confocal configuration, utilizing a 50$\times$ objective and a grating featuring 600 lines per mm. The acquisition comprised 20 accumulations of 2 s each within a single spectral window spanning from 75.12 to 2615.38 cm\textsuperscript{-1}.

\section{Results and Discussions} 
\subsection{Gelation Kinetics}

\begin{figure*}[tb]
    \centering     
    \includegraphics[width=0.9\linewidth]{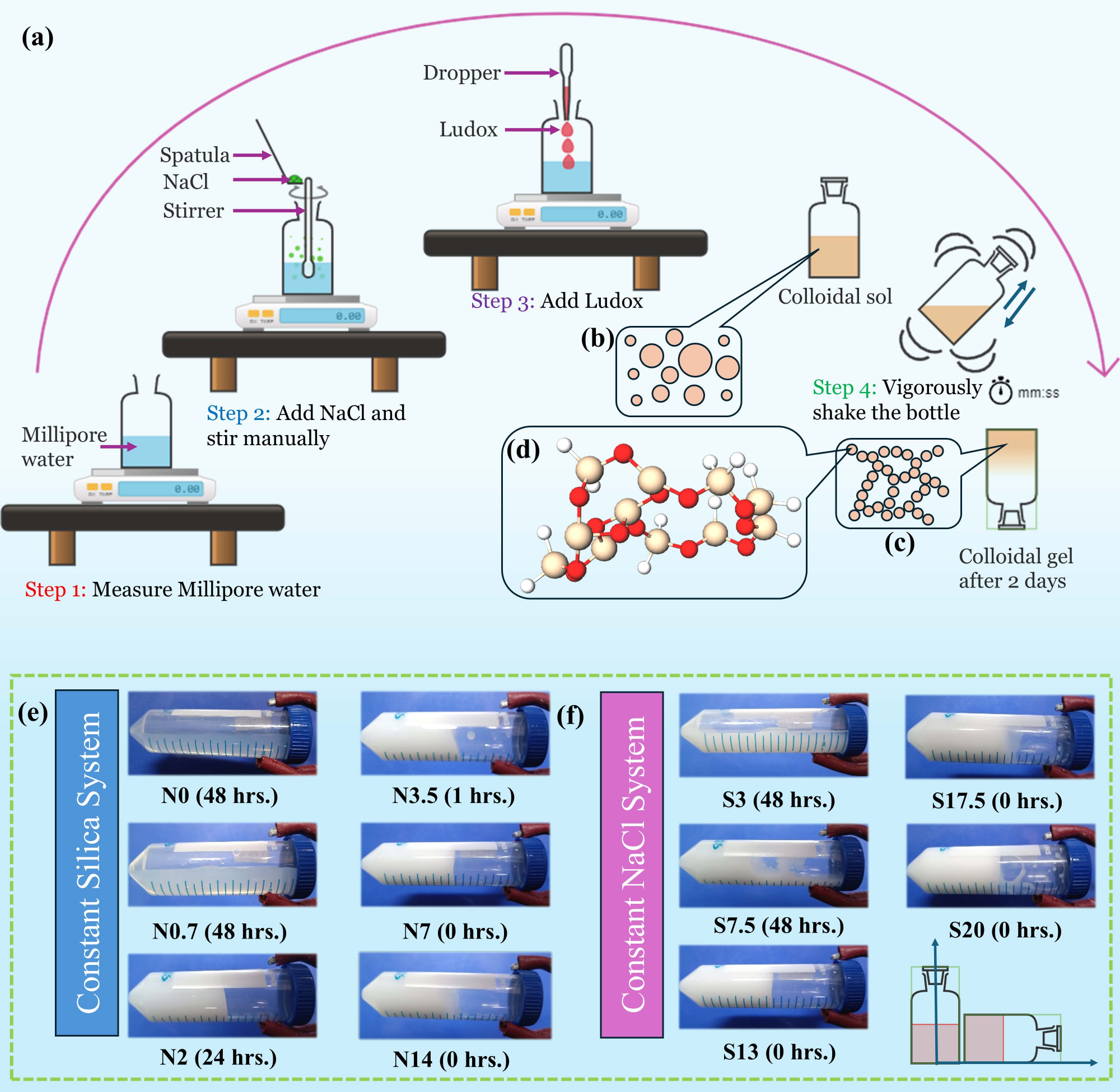}
    \caption{ (a) Schematic illustrating the steps involved in the preparation of a thixotropic colloidal gel of nano-silica: (b) silica sol, (c) gel formation following vigorous shaking, and (d) depiction of the amorphous silica surface. Golden, red, and white spheres represent Si, O, and OH, respectively. The snapshots of (e) constant silica concentration (13 wt.\%) systems with varying NaCl concentration and (f) constant NaCl concentration (7 wt.\%) systems with varying silica concentration at different storage times (values are shown in the bracket). Refer to Table \ref{table:1} for the detailed discussion on different systems.}
    \label{fig:1}
\end{figure*}

 The preparation of colloidal gel (Figure \ref{fig:1} (a-d)) involves mixing a stable stock suspension of negatively charged silica colloidal nanoparticles (Ludox\textsuperscript{\textregistered} TM-40), hereafter referred to as silica/SiO\textsubscript{2}, with an aqueous solution of NaCl. The mixture is vigorously shaken by hand for 5 min. and then left undisturbed for a minimum of 48 hrs. before analysis. Upon adding NaCl to the solution, the Debye length, which suggests the length-scale associated with electrostatic repulsion, decreases. Consequently,  the van der Waals attraction between the particles prevail, leading to their aggregation (see Section S3: SI). Consequently, a space-spanning percolated network structure of silica particles, i.e., gel, forms \cite{Mller2008ShearMaterials}.

The composition of the gel was adjusted in two ways. Firstly, NaCl concentrations were varied while maintaining a silica concentration of approximately 13 wt.\% in the solution. Secondly, the salt concentration is maintained at around 7 wt.\% in the solution, while the silica concentration is varied. For details of formulated gels, refer Table \ref{table:1}. The formulations with 7 wt.\% salt concentration and 13 wt.\% silica, i.e., N7 and S13, are identical. 

\begin{table}[tb]
 \centering
  \caption{Chemical compositions of the gels used in this study}
  \label{table:1}
  \begin{tabular}{c c}
    \hline
   \textbf {Label} & \textbf {Chemical composition} \\
    \hline
    N0 & 0 wt.\% of NaCl + 13wt.\% of silica  \\
    N0.7 & 0.7 wt.\% of NaCl + 13wt.\% of silica\\
    N2 & 2 wt.\% of NaCl + 13wt.\% of silica\\
    N3.5 & 3.5 wt.\% of NaCl + 13wt.\% of silica\\
    N7 & 7 wt.\% of NaCl + 13wt.\% of silica\\
    N14 & 14 wt.\% of NaCl + 13wt.\% of silica\\
    \hline
    S3 & 3 wt.\% of silica + 7wt.\% of NaCl\\
    S7.5 & 7.5 wt.\% of silica + 7wt.\% of NaCl\\
    S13 & 13 wt.\% of silica + 7wt.\% of NaCl\\
    S17.5 & 17.5 wt.\% of silica + 7wt.\% of NaCl\\
    S20 & 20 wt.\% of silica + 7wt.\% of NaCl\\
    \hline
  \end{tabular}
\end{table}

In Figure \ref{fig:1} (e) and (f), we show the snapshots of various systems at different storage times (for more snapshots at different storage times, refer to SI: Section 4, Figure S6, S7). For stable gels, the gelation time varies based on the concentrations of colloidal silica and salt. It is assessed visually by allowing the solution to gel in a transparent bottle. The gel time is determined when the solution ceases to flow upon tilting the container by 90$^\circ$. This method is quick, straightforward, and easy to implement \cite{Sogaard2020FromGels}. The bottle is closely monitored until the solution has fully gelled. We observed that the dispersion exists in different states, i.e., stable gel (N2, N3.5, N7 (S13), N14, S17.5, and S20), viscous liquid (N0 and N0.7), and phase-separated (S3 and S7.5). A significant change in sample turbidity following the addition of NaCl electrolytic solution indicates the rapid formation of aggregates at the micron scale \cite{Kurokawa2015Avalanche-likeGel}. Moreover, under a constant shear rate, the fluid attains a steady state within minutes, which remains stable for hours thereafter \cite{Mller2008ShearMaterials}.

\subsection{Rheological Investigation}\label{Rheological}

A detailed investigation of the rheological properties of thixotropic colloidal silica dispersion is essential to understand their flow behavior, physical aging, and relaxation dynamics. Given that the tribological performance of such systems is strongly influenced by microstructural evolution under both quiescent and flow conditions, rheological analysis provides valuable insight into how the evolution of the material under shear affects lubrication performance. 

To characterize the steady-state flow behavior, shear rate sweep experiments were performed on dispersions with varying concentrations of colloidal silica and NaCl. For dispersions that formed stable gel structures within 48 hours of the storage (N2, N3.5, N7 (S13), N14, S17.5, and S20), samples were first shear melted at a high shear rate of 300 s\textsuperscript{-1} for 1000 s to erase deformation history and establish a reproducible reference state. Whereas, for liquid-like (N0 and N0.7), and phase-separated systems (S3 and S7.5), the samples were first homogenized before being transferred to the rheometer. 

\begin{figure*}[tb]
    \centering      
\includegraphics[width=\linewidth]{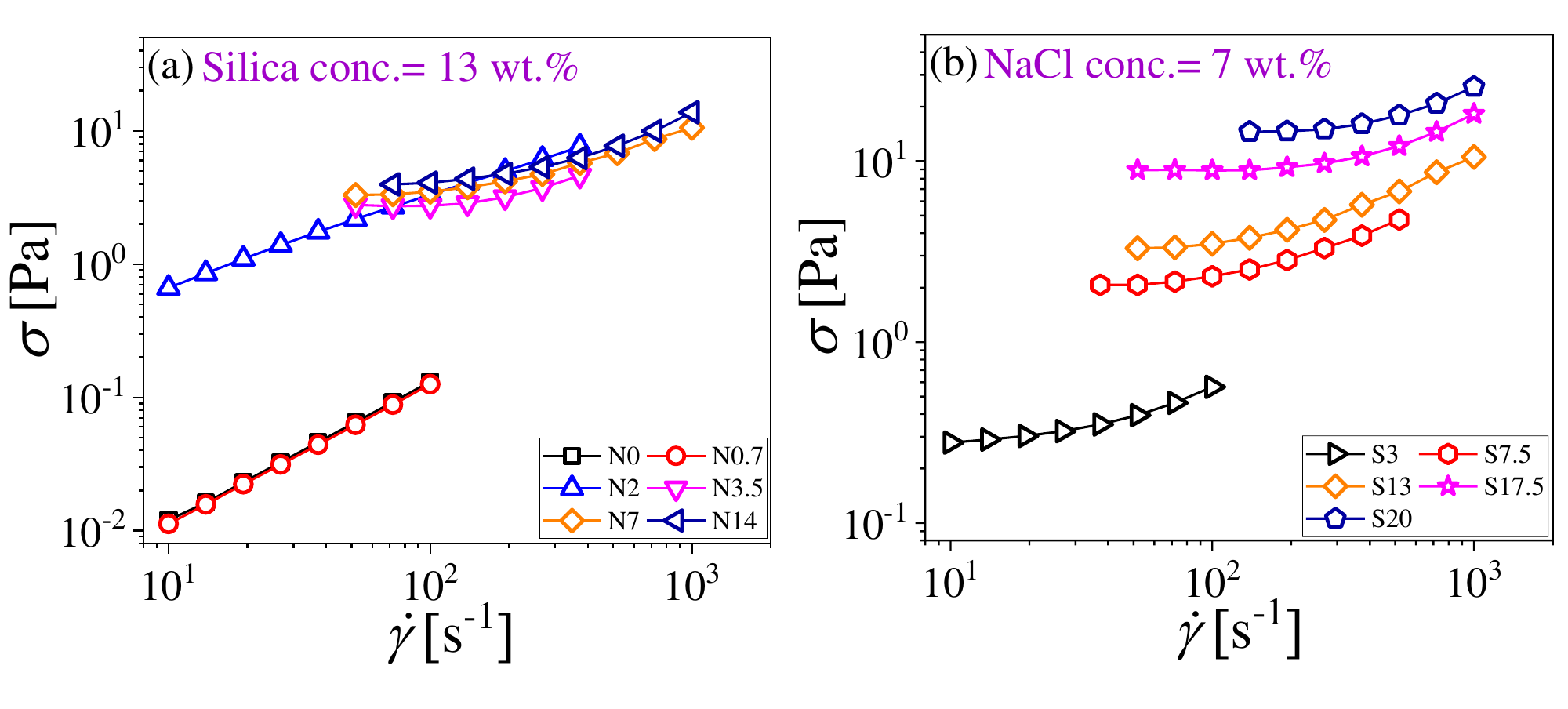}
    \caption{Shear stress is plotted against the shear rate for systems with (a) constant silica particle concentration and (b) constant NaCl concentration. In (a), N0 and N0.7 show Newtonian behavior, while other systems display shear-thinning behavior due to structural breakdown at higher shear rates. In (b), all systems show shear-thinning behavior.} 
    \label{fig:2}
\end{figure*}

The shear rate sweep was performed by progressively decreasing the shear rate from high to low values, maintaining each shear rate for 60 s to allow the system to attain steady-state stress. In Fig. \ref{fig:2} (a), dispersions N0 and N0.7 exhibit Newtonian behavior with shear stress increasing linearly with shear rate, resulting in shear-independent viscosity across the explored range (refer to Section S5.1, Figure S8 in the SI). N0, which lacks NaCl, behaves as a dilute suspension, while N0.7, containing only 0.7 wt.\% NaCl does not form sufficiently large agglomerates to affect flow behavior and also behaves as a Newtonian fluid. Conversely, other systems at constant silica concentration show shear-thinning behavior, arising from the progressive breakdown of internal microstructure and the alignment of the non-Brownian aggregates under flow. Interestingly, the N2 system exhibits higher shear stress at high shear rates compared to N3.5, N7, and N14, despite its lower salt content, likely due to the presence of more numerous but smaller aggregates formed via partial charge screening, leading to enhanced hydrodynamic interactions. Furthermore, N2 lacks a clear stress plateau at low shear rates, which typically characterizes dynamic yield stress in gel-forming systems. Interestingly, systems with higher NaCl concentrations (N3.5, N7, and N14) show the stress plateau, indicating the presence of dynamic yield stress that increases with salt content.

Figure \ref{fig:2} (b) shows the flow curves of systems with constant NaCl concentrations but varying silica particle concentrations. As silica concentration increases, the corresponding shear stress also rises. As previously noted, systems S3 and S7.5 do not form stable gels and exhibit phase separation under quiescent conditions. Despite this, under shear, they demonstrate shear-thinning behavior, indicating the presence of loosely connected agglomerates. However, due to phase separation during quiescent or weak flow states, a reliable dynamic yield stress cannot be defined from their flow curves, despite the appearance of a plateau-like region. In the remaining constant NaCl concentration system (S13, S17.5, and S20), as the silica concentration increases, both the number of agglomerates and the strength of the resulting stable gel network increase, leading to progressively higher values of dynamic yield stress.

It is worth noting that, due to the competition between aging (network formation) and rejuvenation (network destruction) dynamics, thixotropic colloidal gels like silica have been known to show non-monotonic flow behavior \cite{Kumar2025ShearMaterials, Mller2008ShearMaterials}, wherein below a critical shear rate the stress decreases with an increasing shear rate. Since the non-monotonic part of the flow curve is unstable\cite{Sharma2021OnsetDynamics}, the standard stress-controlled rheometer can not be used for measuring the stress in that regime. Importantly, the high shear rates typically encountered in tribological experiments lie outside this unstable regime, making it unlikely to affect lubrication performance. 

During shear rate sweeps, the upper limit of explored shear rates was restricted by the onset of turbulence, while the lower limit was set by the appearance of critical points (flow instabilities) in the agglomerate forming systems or stress fluctuations observed in low-viscosity fluids such as N0 and N0.7.       

\FloatBarrier

\begin{figure*}[tb]
    \centering      
    \includegraphics[width=0.8\linewidth]{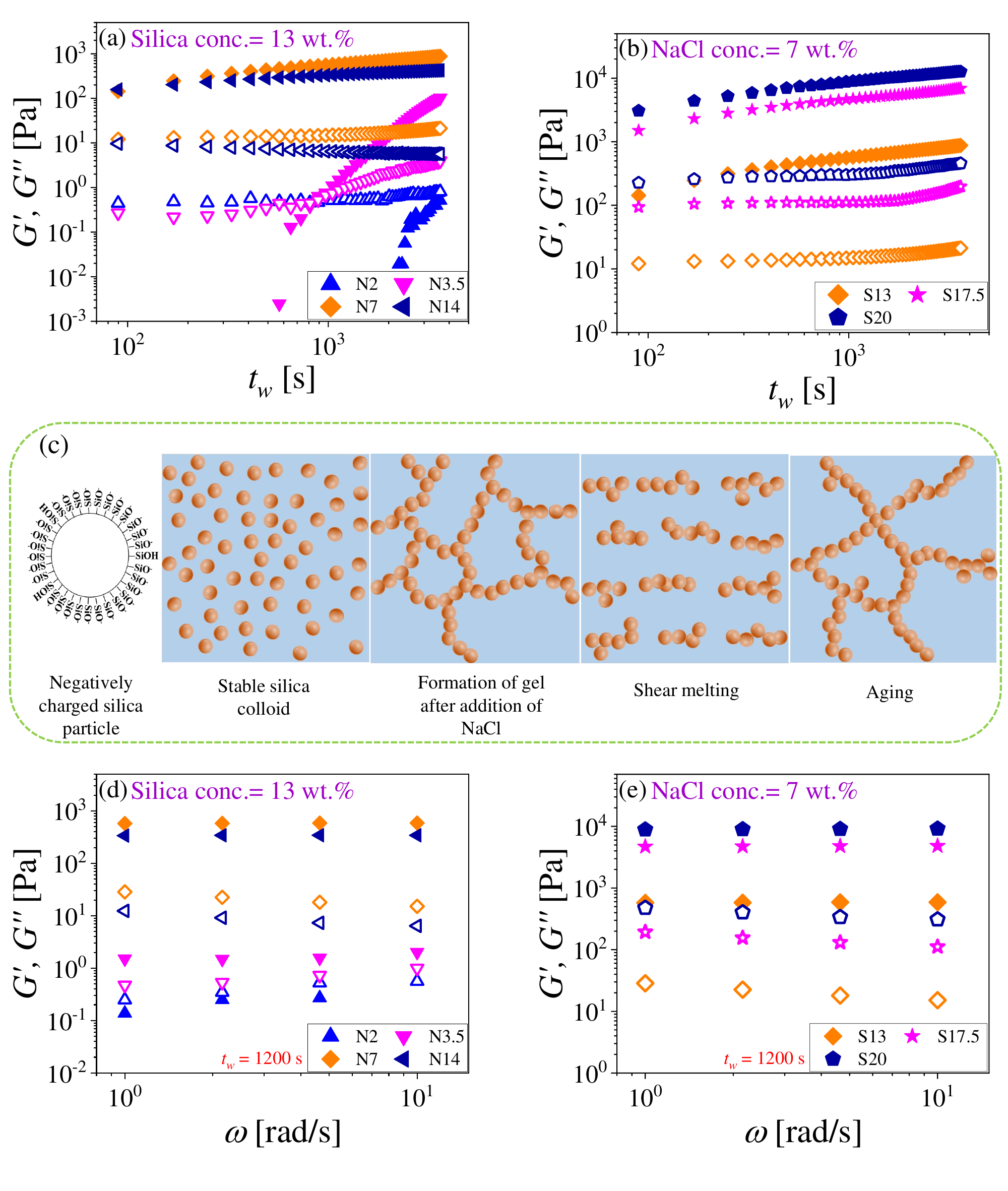}
    \caption{The evolution of elastic modulus $G'$ (solid symbol) and viscous modulus $G''$ (open symbol) with aging time ($t_w$) for (a) constant silica concentration and (b) constant NaCl concentration systems, demonstrating the aging behavior of stable gels. During the aging experiment, the system is subjected to a cyclic frequency sweep for 1 hour subsequent to the shear melting. The data plotted here is for the 0.1 \% strain amplitude and 10 rad/s angular frequency. (c) shows a schematic illustration depicting the gel formation upon NaCl electrolyte addition, shear-induced structural breakdown, and subsequent gel formation during the aging period (red spheres: silica particles). Finally, in (d) and (e), $G'$ (solid symbol) and $G''$ (open symbol) are plotted as a function of angular frequency ($\omega$) for (d) constant silica concentration and (e) constant NaCl concentration systems at $t_w$ = 1200 s.} 
    \label{fig:3}
\end{figure*}

Since N0 and N0.7 systems behave as Newtonian liquids, and S3.5 and S7.5 systems get phase separated, further rheological experiments that explore the aging dynamics of silica dispersion in the quiescent state are only performed on stable gel systems such as N2, N3.5, N7(S13), N14, S17.5, and S20, respectively. To understand the structural recovery and aging behavior, these systems were subjected to the cyclic frequency sweep experiment in the linear visco-elastic domain at $\omega\in{[1, 10]}$ rad/s for 1 hour post-shear rejuvenation. Figures \ref{fig:3} (a) and \ref{fig:3} (b) show the time evolution ($t_w$) of elastic ($G'$) and viscous ($G''$) moduli at an angular frequency of 10 rad/s and 0.1 \% strain amplitude. Here, the waiting time $t_w$ corresponds to the time elapsed since the completion of the shear melting. As can be seen, N2 remains in a liquid-like state ($G''>G'$), while N3.5 transitions into the gel state around 900 s. Interestingly, N7 and N14 systems transition into the gel state instantaneously. Although the elastic moduli of the N14 initially remain higher than N7, the N7 system shows faster-aging dynamics, surpassing the strength of the N14 system within the first 200 s, suggesting faster aging. This behavior is attributed to the optimal balance of inter- and intra-floc interactions in N7, where weaker intra-floc but stronger inter-floc bonding lead to more robust network formation\cite{Shih1990ScalingGels,Metin2014PhaseGel}. Interestingly, compared to the variable NaCl concentration systems, the aging dynamics of the variable silica particle concentration systems show a monotonic increase in gel strength with increasing concentration of silica. As shown in Figure \ref{fig:3} (b), both elastic modulus ($G'$) and viscous modulus ($G''$) increase with an increase in silica concentration. 

The observed aging characteristics and the ability of these stable systems to reform a gel structure after shear rejuvenation are explained schematically in Figure \ref{fig:3} (c). Upon the addition of NaCl, free sodium ions reduce the electrostatic repulsion between the negatively charged silica particles, allowing them to fuse together in the form of non-Brownian aggregates. These aggregates further join together by van der Waals interactions, forming a space-spanning gel network \cite{Kumar2025ShearMaterials, Mller2008ShearMaterials}. When subjected to high shear, these weak van der Waals bonds are disrupted, resulting in flow. Following cessation of shear, bond reformation occurs progressively, driving the system back into the gel state as evident from the time evolution of viscoelastic moduli (Figure \ref{fig:3} (a,b)). This cycle of decrease in viscosity during flow and structure buidup during rest reflects the thixotropic nature of stable silica gels. 

To confirm the gel state, as shown in Figures \ref{fig:3} (d) and (e), frequency sweeps were performed at $t_w$ = 1200 s. Since the frequency sweep experiment takes around 80 s to complete, the effect of aging is very small over the cycle at the waiting time of 1200 s. As can be observed, except for N2, in all other stable-gel forming systems, over the explored frequency range, the elastic modulus ($G'$) remains greater than the viscous modulus ($G''$). Furthermore, $G'$ also remains nearly independent of the angular frequency, confirming the presence of a robust gel structure in these systems. In contrast, for the N2 system, the viscous modulus remains higher than the elastic modulus ($G''> G'$) and both show frequency dependence, suggesting the presence of a weakly structured fluid that has not developed a fully interconnected gel network within the experimental window.

\begin{figure*}[tb]
\centering      
    \includegraphics[width=0.8\linewidth]{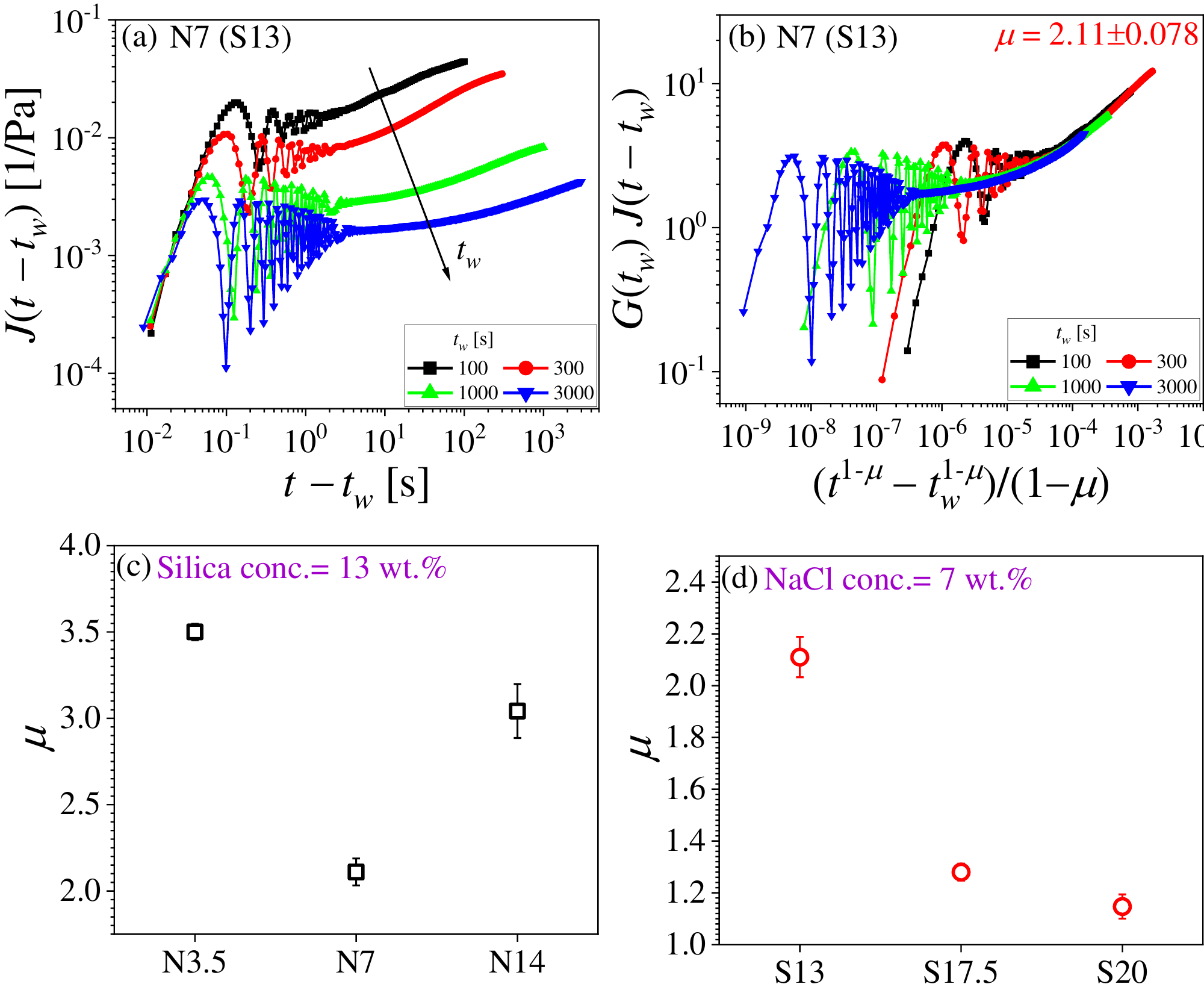}
    \caption{(a) The creep compliance $J(t-t_w)$ is plotted against the creep time $ t-t_w$ at different $t_w$ for N7 (S13). Different symbols assigned for the creep experiment done at different waiting times ($t_w$) are shown in the legend. (b) The scaled creep compliance curves ($G(t_w)J(t-t_w)$) is plotted against the effective time ($(t^{1-\mu} - t_w^{1-\mu} )/(1-\mu)$) for N7(S13). Power law exponents  (\textit{$\mu$}) used for the superposition of the scaled creep compliance curves in the effective time domain is plotted for the (c) constant silica particle concentration and (d) constant NaCl concentration systems.} 
    \label{fig:4}
\end{figure*}

Once the silica systems transition to the gel state, the movement of their microstructural elements becomes increasingly restricted. As a result, the system falls out of thermodynamic equilibrium and seeks to minimize its free energy through the thermal motion of its constituents. This leads to a progressive increase in their relaxation time\cite{Kumar2025ShearMaterials}. Rheologically, this evolution in relaxation time is measured by subjecting the sample to the creep deformation field at different aging times ($t_w$) and measuring the evolution of corresponding creep compliance\cite{Kumar2025ShearMaterials,Kaushal2014LinearMaterials}. The aging process in the colloidal gel systems increases both the elastic modulus ($G'$) and relaxation time with time, resulting in a decrease and downward shift of the induced strain (or creep compliance) when the same magnitude of stress (0.5 Pa) is applied at different $t_w$, as seen in Figure \ref{fig:4} (a) (for the remaining systems, refer to Section S5.2: Figure S9 in SI). Additionally, Figure \ref{fig:4} (a) also reveals that aging impacts not only the magnitude of creep compliance but also decreases its rate of change with creep time ($t-t_w$). Conversely, in the non-aging systems that follow the time translational invariance (TTI) principle, the creep compliance ($J(t-t_w)$) curves remain independent of the waiting time ($t_w$), and the linear viscoelastic behavior of such systems can be explained through the Boltzmann superposition principle, as expressed in Equation 1\cite{ferry1980viscoelastic}

\begin{equation}
    \gamma(t-t_w) = \int_{-\infty}^{t} J(t - t_w) \,\frac{d\sigma}{dt_w} dt_w
\end{equation}

Unlike systems obeying time translational invariance, the creep compliances of silica gels are dependent on both creep time ($t-t_w$) and waiting time ($t_w$). This dependence signifies a departure from TTI\cite{Fielding2000AgingMaterials}, rendering the Boltzmann superposition principle inapplicable in the real-time domain. To address this, Struik, in his 1976 study of aging effects on the relaxation dynamics of glassy polyvinylchloride, introduced the concept of effective time ($\xi$). In this approach, the real-time is scaled by the time-dependent relaxation time through the relation given by Equation 2 \cite{Struik1977PhysicalMaterials, Shukla2020AnalysisGlasses}.

\begin{equation}
\xi (t)= \tau_m\int_{0}^{t} \frac {1}{\tau(t')} \, dt'
\end{equation}
where $\tau_m$ is the constant relaxation time in the effective time domain, and $\mu$=$dln\tau$/$dlnt$ is the power law exponent. Depending on the value of $\mu$, the soft glassy systems are classified as sub-aging ($\mu$$<$1), simple-aging ($\mu$=1), and hyper-aging ($\mu$$>$1) \cite{Shahin2011PredictionMaterials}. Consequently, in the effective time domain, the modified Boltzmann superposition \cite{Shahin2011PredictionMaterials, Kaushal2014LinearMaterials} is expressed as:

\begin{equation}
\gamma(\xi-\xi_w) = \int_{-\infty}^{\xi} J(\xi - \xi_w) \,\frac{d\sigma}{d\xi_w} d\xi_w
\end{equation}
where, $\xi - \xi_w$ is the effective creep time elapsed since the application of stress, and $J(\xi - \xi_w)$ is the creep compliance in the effective time domain. However, to effectively use the effective time approach for analyzing the linear viscoelastic behavior of the soft glassy systems, it's essential to first determine the functional form of the relaxation time, ($\tau(t')$), which is needed for evaluating the integral in Equation 2. Interestingly, aging soft glassy materials often exhibit a power-law relationship between relaxation time and waiting time, expressed as $\tau(t')=A\tau_m^{1-\mu} t'^\mu$ \cite{Shahin2011PredictionMaterials, Kaushal2014LinearMaterials}. Substituting this relationship into Equation 2 yields the following expression of effective time:

\begin{equation} \label{4}
\xi - \xi_w = \frac{\tau_m^\mu}{A}\left[ \frac{t^{1-\mu}-t_w^{1-\mu}}{1-\mu} \right]
\end{equation}

Since in the effective time domain, the relaxation time of the system remains constant, the creep compliance of the system becomes independent of the waiting time ($t_w$), and subsequently, all the scaled creep compliance curves ($G(t_w )J(t-t_w )$) collapse onto a single master curve when plotted against the [$(t^{1-\mu}-t_w^{1-\mu})/(1-\mu)$] for a unique value of $\mu$ as shown in Figure \ref{fig:4} (b) (for the remaining systems, refer to Section S5.2: Figure S10 in SI). Figures \ref{fig:4} (c) and \ref{fig:4} (d) clearly show that all gel-forming silica systems show hyper-aging behavior, characterized by a stronger than linear increase in relaxation time with time. Interestingly, while the magnitude of the relaxation time increases with the strength of the gel structure (large $G'$ at the same $t_w$), its time dependence decreases (smaller $\mu$) due to the greater structural arrest with increased strength. Consequently, we observe that in the case of constant silica concentration systems, the $\mu$ value shows a non-monotonic dependence with respect to the increasing concentration of NaCl, as observed in the case of the evolution of viscoelastic moduli with time in Figure \ref{fig:3} (a). Conversely, for systems with constant NaCl concentration, the $\mu$ value decreases with increasing silica particle concentration. Since viscosity is the product of relaxation time and elastic modulus, this analysis clearly shows the thixotropic nature of the colloidal silica gels, wherein viscosity decreases during the flow (associated with small modulus and relaxation time) and increases at rest. As the thixotropic characteristics of the system depend on the evolution of both elastic modulus and relaxation time, this analysis further shows that while the thixotropic behavior shows a non-monotonic dependence on NaCl concentration, it increases with increasing silica concentration at constant NaCl in stable gel-forming systems.

\subsection{Tribological Study}\label{Tribological}

Metallic tribopairs are often part of machines used in the manufacturing and food processing industries. The leakage of toxic lubricating oil or grease into food during processing poses health hazards. Ensuring consumer health safety while reducing the friction and wear of moving components presents a dual challenge. To address this, we investigated the tribological performance of the formulated gel. In this study we used two tribological methods to evaluate the performance of the lubricant under submerged lubrication conditions. In the first method (M1: \textit{as-prepared lubrication}), the fresh as-prepared lubricant/silica-gel was applied between the tribopair for each test, and the test was conducted for 20 minutes. In the second method (M2: \textit{shear-melted lubrication}), the test initially ran for 20 minute using the as-prepared lubricant, followed by a 20 minute waiting period. Subsequently, the test was repeated for another 20 minutes, and the CoF was recorded during this final test run. The system remained undisturbed and in its original state after the initial 20-minute run. This method is inspired by real-world scenarios where machinery restarts from its previous condition unless disturbed, and the lubricant state remains unchanged during this duration, reflecting common practices in practical lubrication systems. Schematic for both of these test methods is shown in Figure \ref{fig:5} (a-f). 

\subsection{Frictional Behavior Analysis}\label{Friction}

\begin{figure*}[tb]
    \centering         \includegraphics[width=0.8\linewidth]{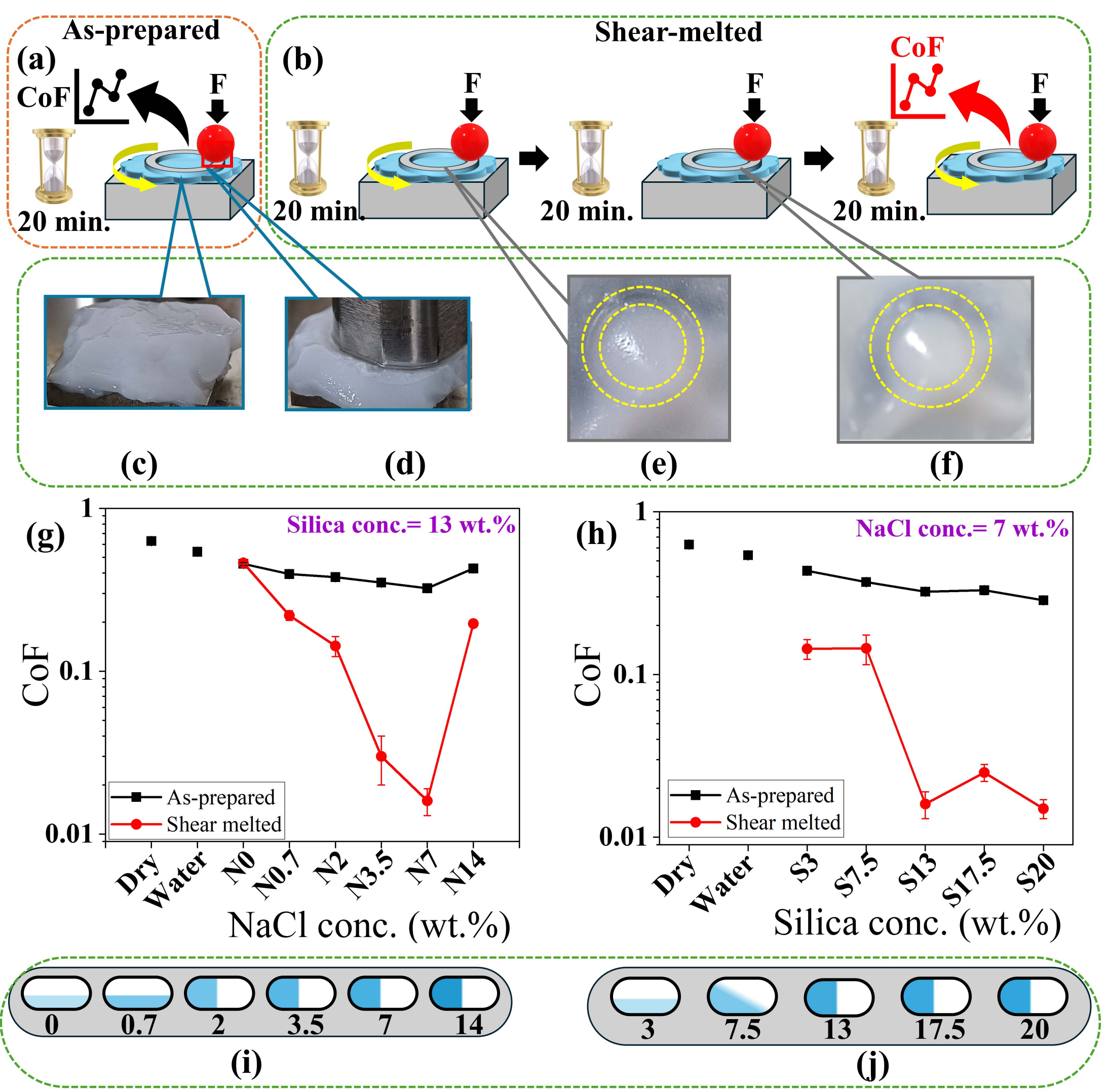}
    \caption{Schematic illustrations depict the process of obtaining CoF using (a) as-prepared lubrication and (b) shear-melted lubrication. Experimental images show (c) the lubricant and (d) the tribo-pair submerged in the lubricant before starting the experiment. The condition of the wear track (highlighted by yellow concentric circles) is shown (e) just after completing the first step of shear-melted lubrication, indicating lubricant removal, and (f) just after the second step (20-minute waiting period), illustrating the rebuilding of the gel structure as the wear track is no longer visible. The mean CoF for as-prepared and shear-melted lubrication is presented (g) at varying salt concentrations (fixed silica concentration of 13 wt.\%) (h) at varying silica concentrations (fixed salt concentration of 7 wt.\%). Error bars represent the standard deviation; for points where error bars are not visible, the standard deviation is within the symbol size. Schematic bars display the flowability of gel lubricants upon tilting the container by 90$^{\circ}$, after 2 days of aging, formulated (i) with varying salt concentrations and (j) with varying silica concentrations. Intensity of color represents viscosity of different gel formulations. Tests were performed at 5 N and 0.04 m/s.}
    \label{fig:5}
\end{figure*}

For fair comparison across various lubrication conditions, we polished the specimens to achieve a uniform surface finish. The average root mean square (RMS) roughness for specimens tested under different lubrication scenarios was 17.664$\pm$3.013 nm (see section S6: SI for details). We compared the frictional performance of silica colloidal gel lubricants with varying concentrations of NaCl (Figure \ref{fig:5}(g)) and silica (Figure \ref{fig:5}(h)). To understand friction variations, we calculated the minimum film thickness using the Hamrock-Dowson formula and determined the Lambda ratio ($\lambda$), which is $<$ 1 for all lubricant formulations, indicating the boundary lubrication regime (see section S7: SI for details). 

Figure \ref{fig:5}(g) shows the mean CoF as a function of silica gel with varying NaCl concentrations. CoF for dry and water-lubricated conditions was also evaluated for comparison. To see the sole effect of NaCl on the friction behaviour, tests were conducted with aqueous NaCl solutions. The results are reported in Figure S11(a) of section S8:SI.  Introducing as-prepared colloidal silica gel without NaCl reduced the CoF by approximately 27\% compared to dry sliding. The CoF further decreased with increasing NaCl concentration, achieving a maximum reduction of about 49\% with 7 wt.\% NaCl (N7). Beyond this concentration, CoF increased, approaching the value for N0, indicating optimal friction performance at N7.
Using the M2 lubrication method, which simulates more practical conditions, yielded interesting results. For N0, the CoF overlapped with the M1 method, as no gel formation occurred without NaCl. However, increasing the salt concentration significantly reduced the CoF, reaching an ultra-low friction state of 0.016 at N7, marking a 97\% reduction compared to dry sliding. For N14, the CoF increased similarly to the M1 method.

Figure \ref{fig:5}(h) illustrates the mean friction coefficient as a function of silica gel with varying silica concentrations. In method M1, with a fixed NaCl concentration of 7 wt.\%, the CoF drops by approximately 31\% for S3, and this reduction continues to increase with higher silica concentrations, reaching about 49\% for S13, which is the same as N7. Further increases in silica concentration yield slightly diminishing reductions in CoF as  $\approx$ 48\% and $\approx$ 54\% for S17.5 and S20 respectively . Under method M2, S3 and S7.5 exhibit similar reductions in CoF, around 77\%, compared to dry sliding. Subsequent increases in silica concentration lead to CoF reductions of approximately 97\% for S13, 96\% for S17.5, and 97\% for S20. The variation of CoF with test duration for the as-prepared and shear-melted states is shown and explained in section S7 of the SI.

To conclude this subsection, NaCl concentration showed a maximum reduction in CoF of 49\% with N7 in method M1, while silica concentration exhibited reductions of approximately 97\% for S13, S17.5, and S20 under method M2. Shear-melted lubricants exhibited stable CoF curves. Overall, S13, S17.5, and S20 showed effective friction reduction and stability, suggesting practical application potential.

\subsection{Wear Resistance Evaluation}\label{Wear}

\begin{figure*}[tb]
    \centering      \includegraphics[width=0.7\linewidth]{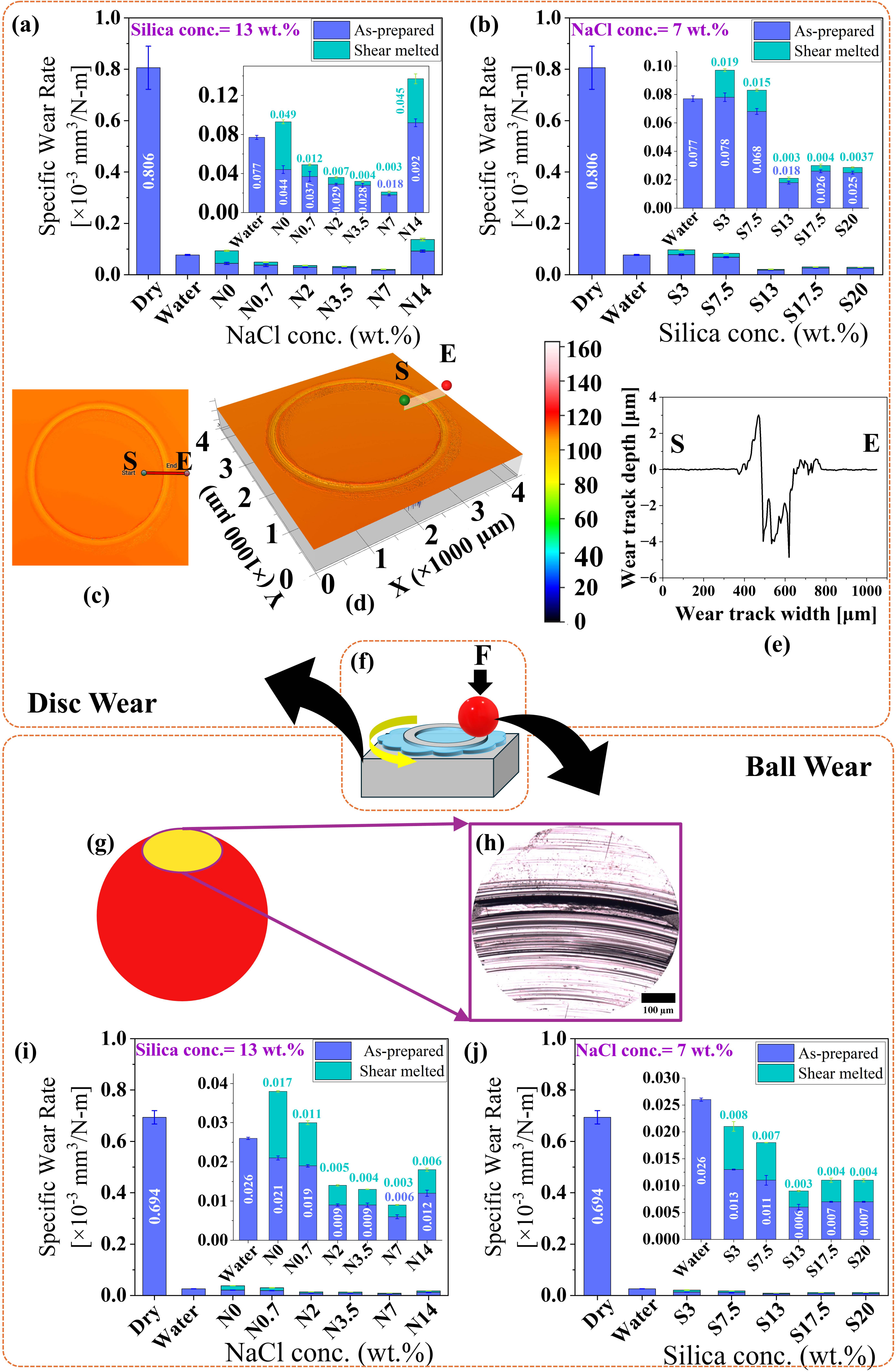}
    \caption{Specific wear rate of the disc for as-prepared and shear-melted colloidal gel lubricant at varying concentrations of (a) salt and (b) silica (c) 2D and (d) 3D images of the wear track, obtained with gel having a 7 wt.\% salt using the shear-melted lubrication method, captured with a 3D optical profilometer. (e) cross-sectional profile of the wear track sliced along S-E; (f) schematic representation of the ball-on-disc lubrication test; (g) schematic illustration of the wear scar on the ball (h) optical microscopic image of the wear scar on the ball tested against the disc lubricated with 7 wt.\% salt formulation using the shear-melted method. Specific wear rate of the ball for as-prepared and shear-melted lubricant at varying concentrations of (i) salt and (j) silica. The inset graph provides a zoomed-in view of the specific wear rate, as the values are too small to be clearly visualized in the main graph. Error bars represent the standard deviation; for columns where error bars are not visible, the standard deviation is within the column size. Tests were performed at 5 N, 0.04 m/s.}
    \label{fig:6}
\end{figure*}

The specific wear rate (SWR) analysis of the disc and ball was conducted under various lubricating conditions using the ISO 18535:2016 method (see SI: section S9 for details). Figure \ref{fig:6}(a) and (b) show the SWR of the disc with as-prepared and shear-melted lubricants at different concentrations of NaCl and silica, respectively.
Figure \ref{fig:6}(a) reveals that in the case of as-prepared silica gel, SWR reduces significantly. Compared to dry sliding, SWR reduces by approximately 94\%, with the reduction increasing to about 98\% at higher NaCl concentrations (N7). However, further salt addition (N14) increases SWR. A similar trend is observed for shear-melted lubrication, shown on top of the stacked column graph. SWR for shear-melted lubrication, calculated after subtracting wear loss from the first step of method M2, shows a maximum reduction of 99.6\% at N7. This suggests the formation of a strong, protective tribolayer beneficial for applications with intermittent sliding. Notably, SWR values for N0 are similar in both conditions, consistent with the shear-independent viscosity shown in Figure \ref{fig:2} (a) resulting from no formation of gel.
Figure \ref{fig:6}(b) shows the SWR variation as a function of silica concentration. A similar trend to that observed with constant silica is evident. Introducing silica reduces SWR by approximately 90\% for S3, with SWR values comparable to those obtained with water, as no gel formation is evident in the flow curve (Figure \ref{fig:2}(b)). The lowest SWR is achieved with S13. Further salt addition does not significantly reduce SWR for S17.5 and S20. This pattern is consistent with shear-melted lubrication results and aligns with the previously discussed friction results. Schematics (Figure \ref{fig:6} (c-h) are explained in the caption of Figure \ref{fig:6}. 

The wear loss of the ball is also significant in this study as it is made of the same material as the disc. Figure \ref{fig:6}(i) shows the SWR as a function of NaCl variation. The SWR values, of the same order ($\times$ 10\textsuperscript{-3}) as the disc, are slightly lower, likely because the disc, with its larger contact area, experiences more uniform and widespread stress, leading to higher wear rates. The trend in SWR reduction is similar to that observed for disc wear. The maximum SWR reduction for the as-prepared condition is 99.1\%, increasing to 99.5\% for shear-melted lubrication, both for the N7 lubricant. A similar trend and explanation apply to the SWR variation with increasing silica concentration. Increasing silica concentration beyond S13 does not significantly reduce SWR (Figure \ref{fig:6}(j)).

\subsection{Discussion}

Our investigation into the frictional and wear performance of colloidal silica gel with thixotropic properties yielded two main findings. First, in as-prepared lubrication, both friction and wear decrease to a certain extent as the sliding friction forces vary with changes in the gel's strength and its viscosity under high contact stress and shear rate. Second, in the shear-melted lubrication method, super-low friction and wear are observed. The key question is: what mechanisms underlie these two methods? We propose that this is due to the synergistic interaction of gel's rheological behavior and surface forces, which govern the deformation and flow of the lubricant under frictional forces. 

Silica sol is electrostatically stabilized in its as-received form and features an extensive amorphous silica network of siloxane bonds (Si$\bm{-}$O$\bm{-}$Si). The surface groups are silanol ($\bm{\equiv}${Si$\bm{-}$OH} ). Geminal and vicinal silanol groups, located a few angstroms apart, form hydrogen bond with water molecules that stabilize the negative surface charge created from deprotonation ($\bm{\equiv}$SiOH $\rightleftharpoons$ $\bm{\equiv}$SiO\textsuperscript{-} + H\textsuperscript{+}). 

As NaCl is added, the counterions present in the electrolyte decrease the surface charge density on the silica nanoparticles and cause the formation of nanoparticle aggregation (flocs) of different sizes through the irreversible Si$\bm{-}$O$\bm{-}$Si bonding between the silica particles. Consequently, these flocs interact through the van der Waals interaction and form the stable gel structure when a sufficient amount of NaCl and silica nanoparticles are present in the dispersion (refer to Figure \ref{fig:1} {(d, e)}) and SI (S2)).

The flow curves in Figure \ref{fig:2} (a) and (b) show shear-thinning behavior for all lubricants except N0 and N0.7, which exhibited Newtonian behavior due to the lack of gel formation. For N0 and N0.7, the counterion concentration in the silica dispersion was insufficient to overcome the electrostatic barrier and consequently, the floc formation did not occur. 
%(\textcolor{red}{remove these lines "collisions between particles in Brownian motion, preventing particle aggregation and resulting in reaction-limited cluster aggregation (RLCA)."}). \textcolor{red}{remove the remaining lines of this paragraph}In contrast, other formulations showed instant to some hours of gel formation due to a high ion concentration, which completely screened electrostatic forces, leading to diffusion-limited cluster aggregation (DLCA). Aggregation resulted in increasing fractal formation, eventually creating a continuous gel network. This explanation also applies when increasing silica concentration while keeping NaCl fixed. A gel with higher particle concentration has closely packed particles with numerous contact areas, where siloxane bonds form \cite{Sogaard2018SilicaAnalysis}, leading to a stronger gel.

At lower concentrations of NaCl or silica, the system either does not form a gel or the gelation time is much longer than the timescale of tribological experiments. Even at higher salt concentrations, like N14, weak gel formation occurred (Figure \ref{fig:3}(a)). 
%\textcolor{red}{due to the faster aggregation of the silica particles into the flocs of the bigger sizes. Subsequently, these flocs combine but form a weaker gel structure due to the formation of a less dense structure}. \textcolor{red}{remove these lines "High salt concentrations neutralize the particles' surface charges, eliminating electrostatic repulsion. With no repulsion to counteract the van der Waals attraction, particles aggregate rapidly and disorderedly, forming dense aggregates or precipitates rather than a gel network. This shifts the system from DLCA to RLCA, where rapid aggregation prevents the fractal growth necessary for gel formation."} 
The viscosity, gel strength, and gelation time vary with silica and salt concentration. In formulations N7, S13, S17.5, and S20, tribological deformation breaks the bonds between flocs, causing the gel to slip between the ball and disc, forming a lubricating film. When sheared gel just comes out of contact zone of ball and disc, the flocs quickly rejoin, reforming the lubricating film and replenishing the wear track during sliding. However, for N2, N3.5, and S7.5, this process occurs partially within the timescale of tribological experiments, resulting in a moderate reduction in friction and wear. Creep tests show that the gel's properties evolve over time, with smaller $\mu$ systems forming gel structures faster and having stronger intermolecular interactions. The gel network also creates pores for water movement \cite{Sogaard2020FromGels}, with dense networks making water passage harder \cite{Sogaard2018SilicaAnalysis}, thus retaining the gel's structure and properties under stress. This stability ensures effective lubrication over time.

The drastic improvement in tribological performance in the shear-melted state can be explained rheologically. Attractive gels under external shear exhibit heterogeneous flows sensitive to pre-shear history \cite{Mller2008ShearMaterials}. Shear melting disrupts the gel network, reduces viscosity, and promotes uniform tribo-film formation at the interface, unlike in the case of as-prepared gel (discussed in the following section). Moderate shear rates cause the gel to reorganize, forming rolling flocs aligned with the sliding direction \cite{Kurokawa2015Avalanche-likeGel}. These flocs rolling along the shear direction reduce friction by minimizing direct contact between sliding surfaces. The gel's ability to reorganize under shear allows it to adapt dynamically to changing conditions, maintaining effective lubrication under varying loads and shear rates.

\subsubsection{Worn Surface Analysis}

\begin{figure*}[tb]
    \centering  \includegraphics[width=0.8\textwidth]{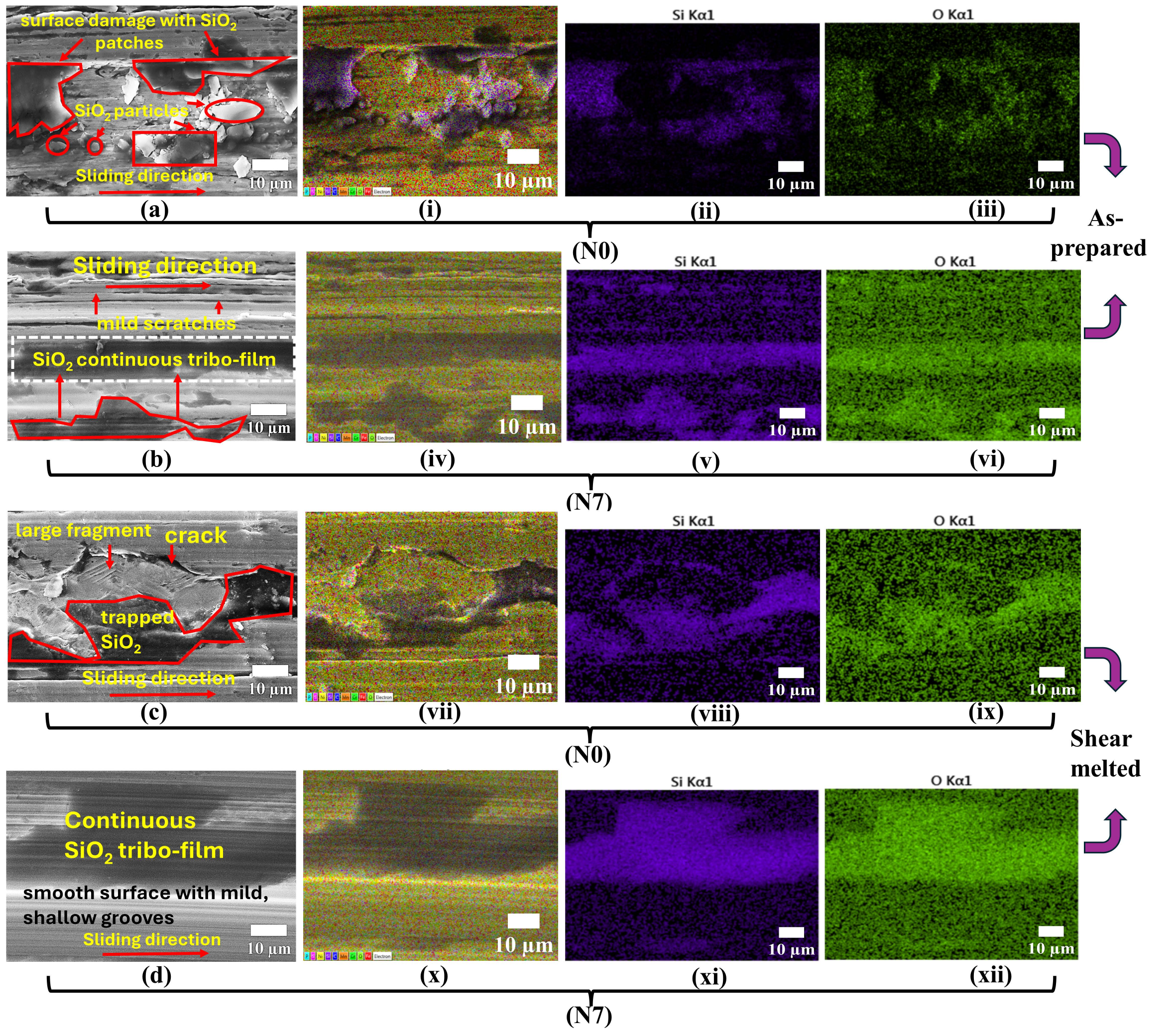}
    \caption{SEM and EDS images of the worn disc surface at different NaCl concentrations in silica: N0 (images a, i, ii, iii) and N7 (images b, iv, v, vi) for the as-prepared lubricant. Images (c, vii, viii, ix) and (d, x, xi, xii) show the worn disc surfaces for N0  and N7, respectively, for shear-melted lubricant.}
    \label{fig:7}
\end{figure*}

To understand the tribological interactions at the steel/steel interface, the worn surface was examined using scanning electron microscopy (SEM) and energy dispersive X-ray spectroscopy (EDS). Figure \ref{fig:7}  shows the SEM and EDS micrographs of the worn disc surface lubricated at different NaCl concentrations in as-prepared and shear-melted lubricant.

The SEM image shown in Figure \ref{fig:7}(a) and (c) shows the worn surface after lubricated by N0 in the as-prepared and shear-melted conditions, respectively. The surface reveals significant damage with large fragment of the steel surface that is about to crack along the sliding direction. This indicates significant mechanical stress and potential failure in the material due to inadequate lubrication. The presence of adhered patches of SiO\textsubscript{2} to the contact areas and entrapment of SiO\textsubscript{2} particles on the worn surface indicates that while silica particles were present, they were not effectively forming a continuous protective film. Instead, they accumulated in damaged areas, unable to provide adequate lubrication. Wear increased due to their inability to withstand shear forces. With an N0 concentration, the silica in the dilute suspension likely failed to reorganize effectively under shear, resulting in poor lubrication performance. EDS images shown in Figure \ref{fig:7}(i-iii and vii-ix) confirm the presence of Si and O on the worn surface. 

The worn surface obtained after using N7 lubricant in the as-prepared and shear-melted conditions is shown in Figures \ref{fig:7} (b) and (d) respectively. It reveals a stable, uniform and continuous tribofilm composed of SiO\textsubscript{2} (confirmed by EDS images (Figure \ref{fig:7} (iv-vi and x-xii)) along the sliding direction, showing the right balance between gel viscosity and stability at optimum NaCl concentration. This tribofilm acts as a barrier between the sliding surfaces, reducing direct metal-to-metal contact and thereby minimizing wear. The surface displays a smooth appearance with only minor scratches to mild and shallow grooves, rather than severe damage, showing the tribofilm’s effectiveness. 

\begin{figure*}[tb]
    \centering  \includegraphics[width=0.8\linewidth]{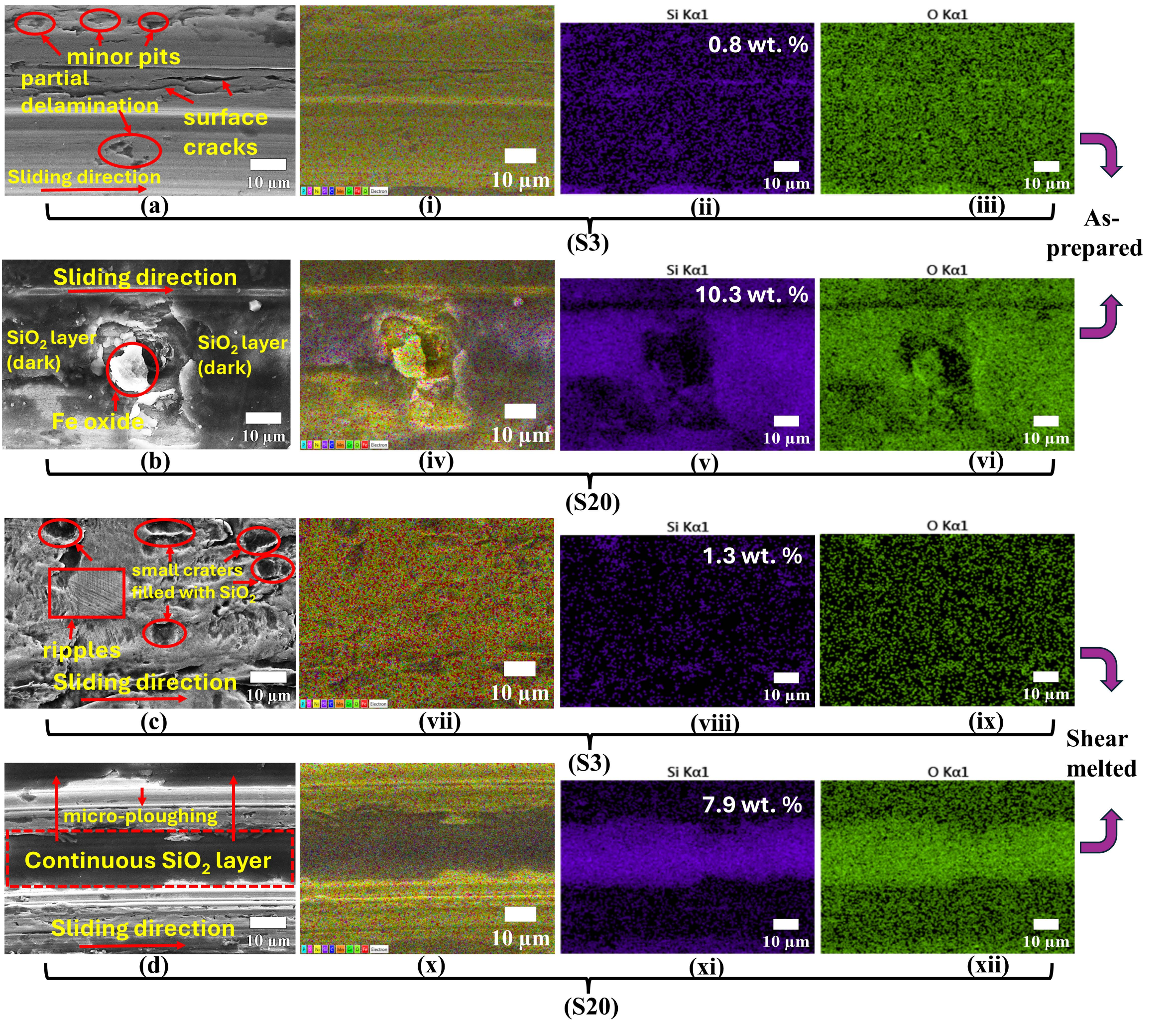}
    \caption{SEM and EDS images of the worn disc surface at different ratios of silica and NaCl: S3 (images a, i, ii, iii) and S20 (images b, iv, v, vi) for the as-prepared lubricant. Images (c, vii, viii, ix) and (d, x, xi, xii) show the worn disc surfaces for S3  and S20, respectively, for shear-melted lubricant. Concentration of Si mentioned in images (ii), (v), (viii) and (xi) is obtained from EDS spectra for the region mapped.}
    \label{fig:8}
\end{figure*}

Figure \ref{fig:8}(a) and (c) show the SEM image of the worn surface after being tested with S3 lubricant in as-prepared and shear-melted conditions, respectively. The surface shows the presence of minor pits and small craters. These pits and craters indicate localized areas of material loss resulting from abrasive wear during sliding. The deposition of SiO\textsubscript{2} in craters suggests that the non-Brownian aggregates released silica particles onto the surface, attempting to provide lubrication and protect the underlying material. However, partial delamination and surface cracks observed on the surface indicate significant mechanical stress during sliding. Also, the presence of ripples on the surface suggests the uneven distribution of lubricating film or localized mechanical stress, leading to surface irregularities., All these surface deformities indicates inability of the lubricating film to adequately protect the surface or distribute the applied load, leading to localized deformation and fracture. Figure \ref{fig:8}(i-iii and vii-ix) confirms the presence of Si and O on the worn surface.

Figure \ref{fig:8}(b) and (d) show the SEM image of the worn surface after being tested with S20 lubricant in as-prepared and shear-melted conditions, respectively. SEM image reveals a consistent and uninterrupted tribofilm composed of SiO\textsubscript{2} (confirmed by EDS images (Figure \ref{fig:8}(iv-vi and x-xii)) along the direction of sliding. This continuous coverage suggests that the colloidal silica gel with the optimum silica concentration effectively formed a stable and uniform protective layer on the steel surface, reducing direct metal-to-metal contact and thus minimizing wear. The surface appears notably smooth, with only mild grooves and evidence of micro ploughing. In conclusion,  SiO\textsubscript{2} tribofilm provided by the silica gel successfully mitigated friction and wear, resulting in minimal surface damage. 

SEM and EDS images of the worn surfaces after testing under dry, water, and lubricant conditions with the remaining formulations (i.e., N0.7, N2, N3.5, N14, S7.5, S17.5) in as-prepared and shear-melted states are illustrated in Figures S14–S17 of Section 10: SI.

\subsubsection{Raman Analysis}

\begin{figure*}[tb]
    \centering  \includegraphics[width=0.8\linewidth]{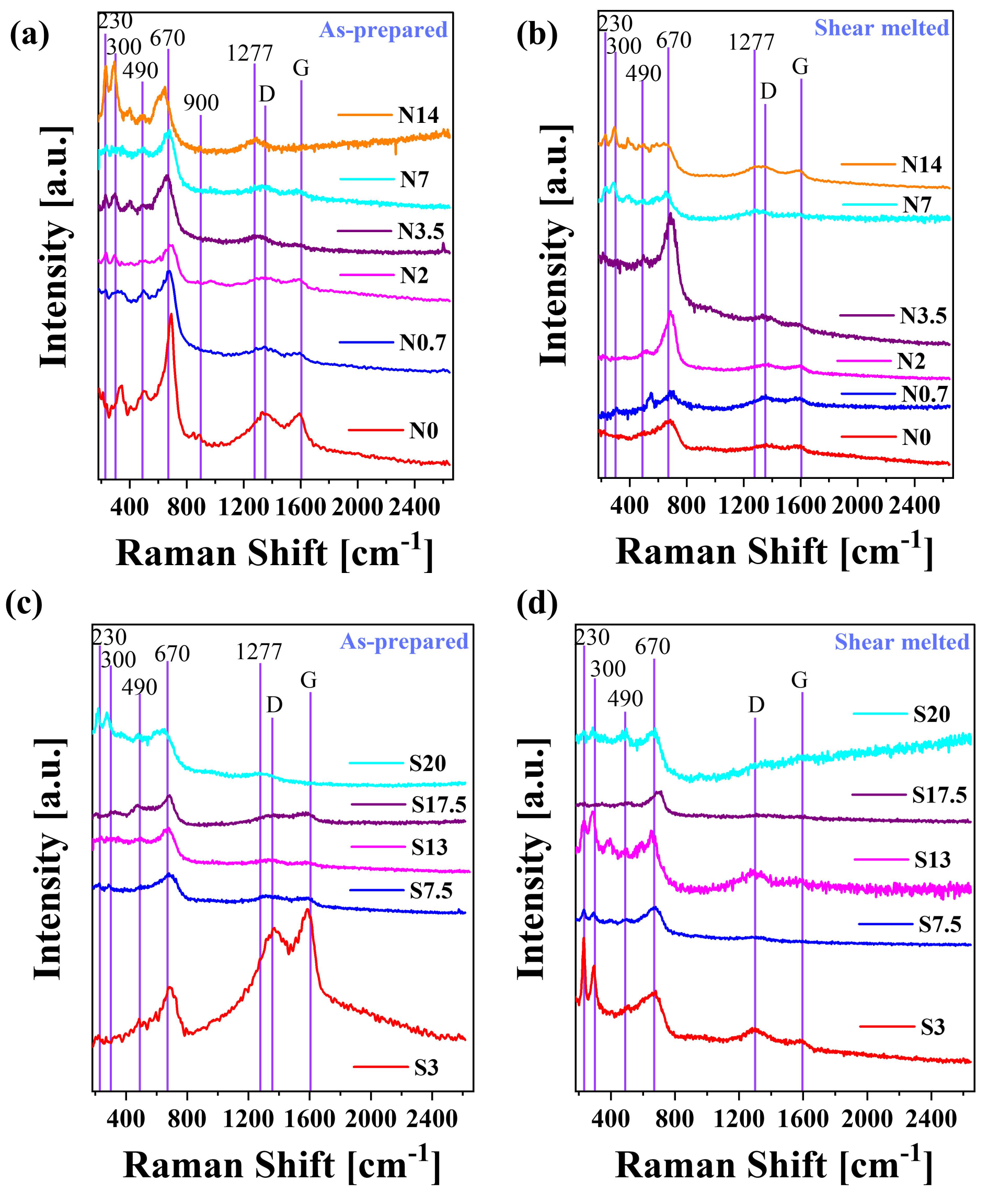}
    \caption{Raman spectra for water-based thixotropic colloidal silica gel at different NaCl (a, b) and silica (c, d) concentrations}
    \label{fig:9}
\end{figure*}

Figure \ref{fig:9}(a) presents the Raman spectra of samples lubricated with the as-prepared lubricant containing varying salt concentrations. All samples exhibited the characteristic silica spectral feature at 490 cm\textsuperscript{-1} \cite{Twu1991RamanSources,Little2008FemtosecondStructure}, corresponding to the symmetric bending modes of four-membered rings \cite{Kadja2017The100C,Khan2017VibrationalMaterials}, and the symmetric stretch vibration of bridging oxygens along the Si$\bm{-}$O$\bm{-}$Si angle bisector with minimal Si motion \cite{Mulder1987THEGEL}. Additionally, SiO\textsubscript{2} peaks were observed with a weak band around 230 cm\textsuperscript{-1} and a relatively stronger band at 300 cm\textsuperscript{-1}, attributed to amorphous silica reflecting the breathing modes of different (Si$\bm{-}$O)\textsubscript{n} ring structures \cite{Mulder1987THEGEL,Geisler2019Real-timeSpectroscopy}. These bands intensified from N0 to N14, suggesting the presence of SiO\textsubscript{2} and indicates that the lubricant did not displace from the wear track and formed a tribofilm on the wear track, reducing friction and wear. A band at approximately 670 cm\textsuperscript{-1}, corresponding to bridging Si$\bm{-}$O$\bm{-}$Si vibrations, was also observed \cite{Borowicz2013RamanDioxide} in all samples. Two intense features, the D peak at approximately 1352 cm\textsuperscript{-1} and the G peak at 1605 cm\textsuperscript{-1} \cite{Guo2021TribologicalNanofluids} can be observed in the case of N0. The D band indicates structural defects in the hexagonal sp\textsuperscript{2} carbon lattice and edge defects, while the G band represents the in-plane vibration of sp\textsuperscript{2} carbon atoms \cite{Autthawong2020Cost-effectiveAnodes}. This suggests high wear, as without any salt, no gel structure forms. As salt concentration increases from N0 to N14, the intensity of the D and G bands decreases, indicating reduced disorder and fewer local defects. For N7 and N14, a very weak Raman signal indicates minimal carbon wear. In N14, the G band disappears, and the D band shifts slightly to 1277 cm\textsuperscript{-1}, corresponding to the silica symmetric O$\bm{-}$Si$\bm{-}$O stretch ($\approx$1200 cm\textsuperscript{-1}) \cite{Lagstrom2015SurfaceDensity}. This shift may be attributed to shearing actions at the mating interface \cite{Ding2018TheAdditives}. 

Similar Raman spectra were observed for shear-melted lubricants with varying salt concentrations (Figure \ref{fig:9}(b)). In these samples, the D and G bands were less intense compared to those in Figure \ref{fig:9}(a), with a very weak D band and the disappearance of the G band in N7. This indicates reduced wear, consistent with the wear results shown in Figure \ref{fig:6}. The spectral peaks for as-prepared lubricants with varying silica concentrations (Figure \ref{fig:9}(c)) are similar to those in Figure \ref{fig:9}(a). For S3, with low silica concentration, the SiO\textsubscript{2} peaks at 230 and 300 cm\textsuperscript{-1} disappear, and the D and G bands intensify, indicating surface wear, as shown in Figure \ref{fig:6}. As silica concentration increases, these SiO\textsubscript{2} peaks reappear, and the D and G bands decrease in intensity, suggesting stronger gel formation and silica retention at the wear track. The Raman spectra for worn samples lubricated with shear-melted lubricant with varying silica concentrations (Figure \ref{fig:9}(d)) are similar to those in Figure \ref{fig:9}(b). In these samples, most peaks remain the same except for a slight shift of the D and G bands to ~1300 cm\textsuperscript{-1} and 1592 cm\textsuperscript{-1}, which almost disappear in S17.5 and S20, indicating a stable and continuous silica tribofilm.

\subsubsection{Investigation into Lubrication Mechanism}

\begin{figure*}[tb]
    \centering  \includegraphics[width=0.86\linewidth]{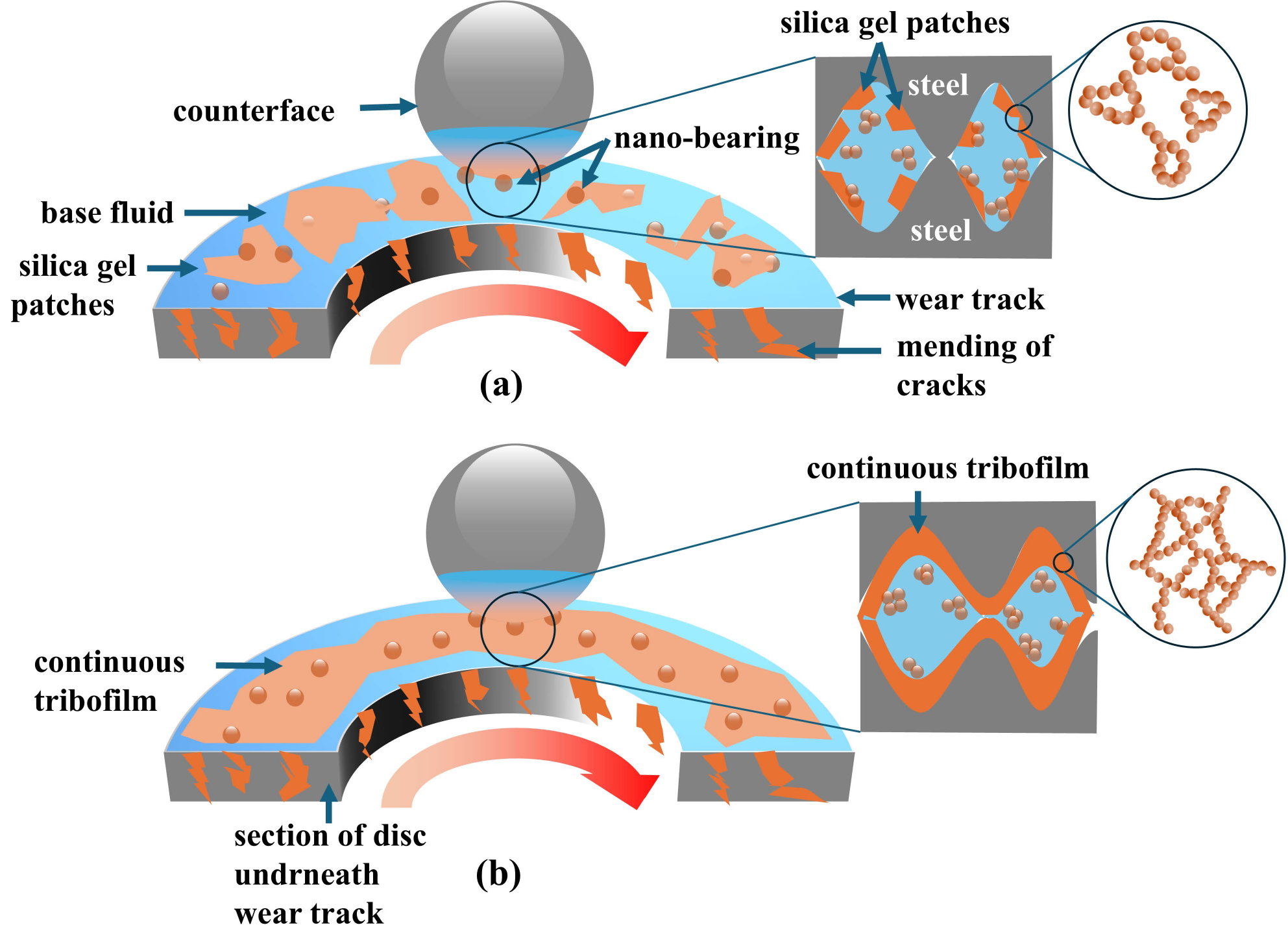}
    \caption{Schematic representation of the lubrication mechanism in water-based thixotropic colloidal silica gel (a) formation of SiO\textsubscript{2} patches (b) fully grown SiO\textsubscript{2} tribofilm }
    \label{fig:10}
\end{figure*}

Figure \ref{fig:10} depicts a plausible lubrication mechanism for water-based thixotropic colloidal silica gel, based on friction and wear performance. We propose that lubrication results from a synergistic combination of gelation kinetics, the mending effect, tribo-film evolution and the nano-bearing effect \cite{Hao2022FrictionAdditive} at the interface.

When the as-prepared lubricant is introduced between the rotating disc and ball, micron-scale SiO\textsubscript{2} flocs \cite{Kurokawa2015Avalanche-likeGel,Bailey1989MICROSTRUCTURALSOLS} are dragged into the interface with the base fluid. The shearing action deforms Si$\bm{-}$O$\bm{-}$Si bonds \cite{Wen2016AtomicField}. Due to its low viscosity, the lubricant fills surface irregularities and cracks, as small as 15 $\mu$m \cite{Sogaard2020FromGels}, generated during sliding. Once the cracks are filled, the SiO\textsubscript{2} flocs compact and form SiO\textsubscript{2} patches (Figure \ref{fig:10}(a)) which subsequently grow on the surface (Figure \ref{fig:10}(b)) and physically adhere to the surface asperities, forming a protective tribo-film \cite{Guo2018EnhancedAdditive}. Due to the microscale surface roughness, the local contact pressure at the asperities is higher and not uniform compared to the Hertzian contact pressure. In these areas, water alone cannot separate the surfaces, but the colloidal silica gel layer, with its comparatively high local viscosity \cite{Ootani2020Self-FormedEnvironment}, can effectively provide separation. Also, N7 or S13 exhibits stronger gel formation and smaller gelation time due to faster gelation kinetics. This is attributed to the optimal salt concentration, which effectively screens electrostatic repulsions between the silica particles, thereby reducing the Debye screening length. This allows the gel to reform the structure quickly post shear and provide a protective interfacial film continuously. This restricts the tribosystem from running in a starved lubrication system. When the salt or silica concentration is too low, it is difficult to form a protective tribo-film, leading to direct contact between the sliding steel pairs and resulting in a high friction coefficient. Conversely, when the additive concentration is too high, SiO\textsubscript{2} nanoparticles tend to aggregate into large particles, causing an abrasive effect and obstructing the sliding, which also leads to high friction \cite{Liu2020PreparationAdditive}. Localized rolling of small clusters of spherical SiO\textsubscript{2} nanoparticles embed onto the disk surface, acting as nano-bearings \cite{Ding2018TheAdditives}, is hypothesized. In low-viscosity zones and confined asperity valleys, such clusters can undergo partial rotational motion. This cluster-mediated nano-bearing effect, although localized, is believed to assist in reducing interfacial friction through rolling-sliding interactions. 

\section{Conclusion}\label{sec4}

To conclude, this study highlights the potential of water-based thixotropic colloidal gel of SiO\textsubscript{2} nanoparticles as a lubricant for metallic tribosystems. It demonstrates how the synergistic effects of rheology and tribology can achieve super-low friction and wear. By optimizing the gel's rheological properties i.e., viscosity, gel strength, aging dynamics, and yielding behavior, we fine-tuned its efficacy. The N7 gel formulation demonstrates outstanding tribological performance, achieving up to 97.46\% (from 0.63 to 0.016) and 97.04\% (from 0.541 to 0.016) reduction in the friction coefficient compared to dry and water lubrication, respectively, under shear-melted conditions. Similarly, it delivers a remarkable 99.62\% (from 0.806 to 0.003 $\times$ 10\textsuperscript{-3} mm\textsuperscript{3}/N-m) and 96.10\% (from 0.077 to 0.003 $\times$ 10\textsuperscript{-3} mm\textsuperscript{3}/N-m) decrease in specific wear rate. The exceptionally low CoF and wear achieved are attributed to the thixotropic and chemically robust nature of gel, formed through van der Waals interaction between the flocs, which provides self-repairing properties, continuous tribo-film formation and nano-bearing effect. These factors help maintain the lubricant in the interfacial contact zone, even under high contact pressure and large shear forces, allowing the tribosystem to operate in a fully submerged lubrication state rather than the lubrication-starved condition often observed with greasy lubricants. This study lays the groundwork for future research aimed at understanding the lubrication mechanisms on an atomic scale, potentially improving the wear and friction performance of metallic tribosystems by fine-tuning microscopic contact mechanics and gel structures. Also, its simple, non-toxic composition and scalable preparation process highlight its potential for use in food-grade and environmentally sensitive applications. This work can pave way for the development of cost-effective and eco-friendly lubricants.

\section*{CRediT authorship contribution statement}
\textbf{AK}: Methodology, Investigation, Data curation, Formal analysis, Writing-original draft. \textbf{VK}: Methodology, Investigation, Data curation, Formal analysis, Writing–original draft. \textbf{YMJ}: Conceptualization, Funding acquisition, Supervision, Writing–review and editing. \textbf{MKS}: Conceptualization, Funding acquisition, Supervision, Writing–review and editing. 

\section*{Declarations}
The authors declare no competing financial interest.

\section*{Acknowledgement}
YMJ acknowledges financial support from the Science and Engineering Research Board,
Government of India (Grant numbers: CRG/2022/004868 and JCB/2022/000040).

\section*{Supporting Information Available}
Schematic of experimental set-up for tribological tests and ball-on-disc assembly associated with the set-up, change in CoF with entrainment speed at load of 1N, 2N, 3N, 5N and 10N. CoF variation at a sliding speed of 0.04 m/s under 1N, 2N, 3N, 5N and 10N load using N7(S13) lubricant, illustration of ion interaction with a silica surface and its effect on the Debye length, snapshots of different systems at varying storage time, viscosity vs. shear rate, evolution of creep compliance $J(t-t_w)$ against creep time $(t-t_w)$ for different $t_w$, time-aging time superposition for stable gels, surface characterization of unworn disc, description of the methodology adopted for evaluation of lubrication regime, friction behavior of aqueous NaCl solutions, Specific wear rate of disc at sliding speed of 0.04 m/s and at load of 1N, 2N, 3N, 5N and 10N, variation of CoF over time for lubricants with varying NaCl concentra-
tions and fixed silica concentration of 13 wt.\% in (a) as-prepared state (b) shear-melted state. Variation of CoF over time for lubricants with different silica concentrations and fixed NaCl concentration of 7 wt.\% in (a) as-prepared state and (b) shear-melted state. description of the methodology adopted for specific wear rate calculation, SEM and EDS micrographs of wear mechanism for dry, water, N0.7, N2, N3.5, N14, S7.5, S17.5 in as-prapared and shear melted lubrication. 

\section*{Data availability}
Data will be made available on reasonable request.

% \appendix*
% \input{sections/appendix1.tex}

\bibliography{References}
\bibliographystyle{apsrev4-1}
\end{document}